\documentclass[twocolumn,aps,prl,groupedaddress,showpacs]{revtex4}
\usepackage[latin9]{inputenc}
\usepackage{color}
\usepackage{amsmath}
\usepackage{amssymb}
\usepackage{graphicx}
\usepackage[unicode=true,pdfusetitle,
 bookmarks=true,bookmarksnumbered=false,bookmarksopen=false,
 breaklinks=false,pdfborder={0 0 1},backref=false,colorlinks=true]
 {hyperref}
\usepackage{breakurl}

\makeatletter

\newcommand{\lyxmathsym}[1]{\ifmmode\begingroup\def\b@ld{bold}
  \text{\ifx\math@version\b@ld\bfseries\fi#1}\endgroup\else#1\fi}

\providecommand{\tabularnewline}{\\}

\@ifundefined{textcolor}{}
{%
 \definecolor{BLACK}{gray}{0}
 \definecolor{WHITE}{gray}{1}
 \definecolor{RED}{rgb}{1,0,0}
 \definecolor{GREEN}{rgb}{0,1,0}
 \definecolor{BLUE}{rgb}{0,0,1}
 \definecolor{CYAN}{cmyk}{1,0,0,0}
 \definecolor{MAGENTA}{cmyk}{0,1,0,0}
 \definecolor{YELLOW}{cmyk}{0,0,1,0}
}

\usepackage{color}

\makeatother

\begin{document}

\title{Quantum Superinductor with Tunable Non-Linearity }

\author{M.T. Bell,$^{1}$ I.A. Sadovskyy,$^{1}$ L.B. Ioffe,$^{1,2}$ A.Yu.
Kitaev,$^{3}$ and M.E. Gershenson$^{1}$}

\affiliation{$^{1}$Department of Physics and Astronomy, Rutgers University, Piscataway,
New Jersey 08854, USA}

\affiliation{$^{2}$LPTHE, CNRS UMR 7589, 4 place Jussieu, 75252 Paris, France}

\affiliation{$^{3}$Caltech, Institute for Quantum Information, Pasadena, California
91125, USA}
\begin{abstract}
We report on the realization of a superinductor, a dissipationless
element whose microwave impedance greatly exceeds the resistance quantum
$R_{{\rm Q}}$. The design of the superinductor, implemented as a
ladder of nanoscale Josephson junctions, enables tuning of the inductance
and its nonlinearity by a weak magnetic field. The Rabi decay time
of the superinductor-based qubit exceeds 1\,$\mu\textrm{s}$. The
high kinetic inductance and strong nonlinearity offer new types of
functionality, including the development of qubits protected from
both flux and charge noises, fault tolerant quantum computing, and
high-impedance isolation for electrical current standards based on
Bloch oscillations. 
\end{abstract}

\pacs{74.50.+r, 74.81.Fa, 85.25.Am}

\maketitle
Superinductors are desired for the implementation of the electrical
current standards based on Bloch oscillations \cite{Likharev:1985,Averin:1991},
protection of Josephson qubits from the charge noise \cite{Manucharyan:2009,ManucharyanThesis:2012},
and fault tolerant quantum computation \cite{Kitaev:2006,Doucot:2012}.
The realization of superinductors poses a challenge. Indeed, the \textquotedblleft{}geometrical\textquotedblright{}
inductance of a wire loop is accompanied by a sizable parasitic capacitance,
and the loop impedance $Z$ does not exceed $\alpha R_{\textrm{Q}}$
\cite{Feynman:1964}, where $\alpha\lyxmathsym{ }=\lyxmathsym{ }1/137$
is the fine structure constant and $R_{\textrm{Q}}=h/4e^{2}$ is the
resistance quantum. This limitation does not apply to superconducting
circuits whose kinetic inductance $L_{\textrm{K}}$ is associated
with the inertia of the Cooper pair condensate \cite{Tinkham:1996}. 

The kinetic inductance of a Josephson junction, $L_{\textrm{K}}=(\Phi_{0}/2\pi)^{2}/E_{\textrm{J}}$,
scales inversely with its Josephson energy~$E_{\textrm{J}}$ \cite{Tinkham:1996}
($\Phi_{0}=h/2e$ is the flux quantum). The kinetic inductance can
be increased by reducing the in-plane junction dimensions and, thus,
$E_{\textrm{J}}$. However, this resource is limited: with shrinking
the junction size, the charging energy $E_{\textrm{C}}=e^{2}/2C$
($C$ is the junction capacitance) increases and the phase-slip rate
$\propto\exp[-(8E_{\textrm{J}}/E_{\textrm{C}})^{1/2}]$ \cite{Tinkham:1996}
grows exponentially, which leads to decoherence. Small Josephson energy
(i.e., large kinetic inductance) can be realized in chains of dc SQUIDs
frustrated by the magnetic field \cite{Watanabe:2004,Corlevi:2006}.
However, the phase-slip rate increases greatly with frustration, and
the chains do not provide good isolation from the environment. For
the linear chains of Josephson junctions with $E_{\textrm{J}}/E_{\textrm{C}}\gg1$,
relatively large values of $L_{\textrm{K}}$ (up to 300\,nH \cite{Manucharyan:2009})
have been realized in the phase-slip-free regime; further increase
of the impedance of these chains is hindered by the growth of their
parasitic capacitance. Also, the linear chains, as well as the nanoscale
superconducting wires with a kinetic inductance of $\sim\lyxmathsym{ }10$\,nH/$\mu$m
\cite{Annunziata:2010,Manucharyan:2010}, are essentially linear elements
whose inductance is not readily tunable (unless large currents are
applied).

We propose a novel superinductor design that has several interesting
features. This circuit can be continuously tuned by a weak magnetic
field between the regimes characterized by a low linear inductance
and a very large nonlinear inductance. Importantly, the large impedance
$Z\gg R_{\textrm{Q}}$ is realized when the decoherence processes
associated with phase slips are fully suppressed. This combination
of strong nonlinearity and low decoherence rate is an asset for the
development of high-performance superconducting qubits and controllable
coupling between qubits.

\begin{figure*}[t]
\includegraphics[width=0.8\textwidth]{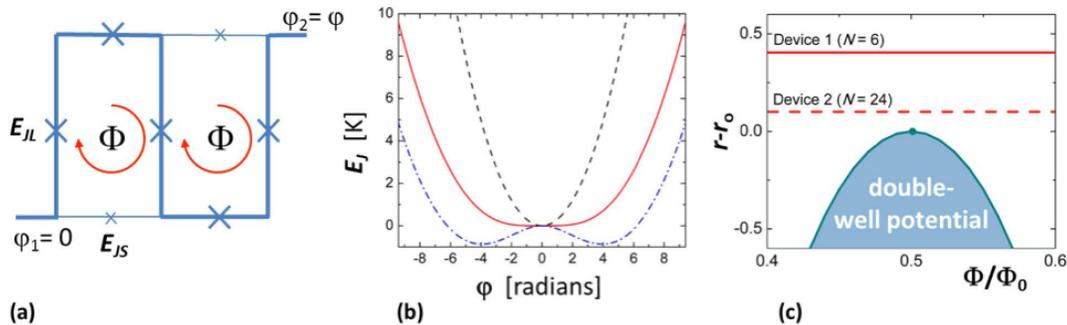} \caption{(color online) Superinductor unit cells, potential energy, and phase
diagram. (a) The unit cells of the tested device include \textquotedblleft{}small\textquotedblright{}
and \textquotedblleft{}large\textquotedblright{} Josephson junctions
with Josephson energies $E_{\textrm{JS}}$ and $E_{\textrm{JL}}$,
respectively. The \textquotedblleft{}backbone\textquotedblright{}
of the superinductor is shown as a bold line; the adjacent cells share
large Josephson junctions. The cells are threaded by the same magnetic
flux $\Phi$; the phase difference across the device is $\varphi_{2}-\varphi_{1}=\varphi$.
(b) The Josephson energy $E_{\textrm{J}}(\varphi)$ of a six-cell
ladder calculated within the quasiclassical approximation at $\Phi\lyxmathsym{ }=\lyxmathsym{ }0$
(dashed curve) and $\Phi\lyxmathsym{ }=\lyxmathsym{ }\Phi_{0}/2$
(solid and dash-dotted curves); $E_{\textrm{JS}}=3.5$\,K for all
three curves, $E_{\textrm{JL}}=16.8$\,K for dashed and solid curves
and $E_{\textrm{JL}}=14$\,K for the dash-dotted curve. For the optimal
value of the ratio $E_{\textrm{JL}}/E_{\textrm{JS}}$, denoted as
$r_{\textrm{o}}$, the dependence $E_{\textrm{J}}(\varphi)$ becomes
flat near $\varphi\lyxmathsym{ }=\lyxmathsym{ }0$ at $\Phi\lyxmathsym{ }=\lyxmathsym{ }\Phi_{0}/2$
(solid curve). For smaller values of $E_{\textrm{JL}}/E_{\textrm{JS}}$,
a double well potential is realized near full frustration. (c) \textquotedblleft{}Phase
diagram\textquotedblright{} of the ladders on the $r$-$\Phi$ plane.
The values of $r_{\textrm{o}}$ are 4.1 and 4.5 for the ladders with
$N=6$ and 24, respectively. The values of $r-r_{\textrm{o}}$ for
the studied devices are shown as horizontal lines. }

\label{fig:setup} 
\end{figure*}

The studied circuit is a \textquotedblleft{}ladder\textquotedblright{}
of nanoscale Josephson junctions frustrated by the magnetic flux $\Phi$
{[}Fig. \ref{fig:setup}(a){]}. Each unit cell of the ladder represents
an asymmetric dc-SQUID with a single \textquotedblleft{}small\textquotedblright{}
junction with the Josephson energy $E_{\textrm{JS}}$ in one arm and
three \textquotedblleft{}large\textquotedblright{} Josephson junctions
with the Josephson energy~$E_{\textrm{JL}}$ in the other arm (the
in-plane dimensions of both types of tunnel junctions do not exceed
$0.3\times0.3\,\textrm{\ensuremath{\mu}m}^{2}$). The adjacent cells
are coupled via one large junction; the Josephson energy of the system,
$E_{\textrm{J}}(\varphi)$, remains an even function of the phase
difference $\varphi$ across the ladder at any value of the flux (the
benefits of this symmetry are discussed below).

For the Josephson junctions in the ladder \textquotedblleft{}backbone,\textquotedblright{}
the Josephson energy is two orders of magnitude larger than the charging
energy (the junction parameters are summarized in Table~\ref{tab:data}).
In this case quantum fluctuations of the phase across individual junctions
are small, and the dependence $E_{\textrm{J}}(\varphi)$ {[}see Fig.\ref{fig:setup}(b){]}
can be obtained from the classical computation that minimizes~$E_{\textrm{J}}$
with respect to the phases of superconducting islands at a fixed~$\varphi$.
At zero frustration ($\Phi=0$) the energy $E_{\textrm{J}}(\varphi)$
is approximately parabolic over the relevant range of~$\varphi$.
With an increase of the magnetic field, the curvature of $E_{\textrm{J}}(\varphi)$
near $\varphi=0$ decreases. Provided the ratio of the Josephson energies
for large and small junctions, $r=E_{\textrm{JL}}/E_{\textrm{JS}}$,
is not too large, the curvature $\partial^{2}E_{\textrm{J}}(\varphi)/\partial\varphi^{2}|_{\varphi=0}$
vanishes at some critical flux $\Phi_{\textrm{c}}$. At this frustration
the potential is strongly anharmonic (approximately quartic) and the
kinetic inductance, which is inversely proportional to the curvature
of the potential $L{}_{\textrm{K}}\propto[\partial^{2}E_{\textrm{J}}(\varphi)/\partial\varphi^{2}]^{-1}$
\cite{Tinkham:1996} in its minimum, diverges.

The optimal regime of operation for the studied superinductor and
superinductor-based qubits is realized when the curvature $\partial^{2}E_{\textrm{J}}(\varphi)/\partial\varphi^{2}|_{\varphi=0}$
vanishes exactly at full frustration ($\Phi=\Phi_{0}/2$). Indeed,
because the energy is an even function of the flux at full frustration,
the device in this regime is insensitive in the first order to the
flux noise \cite{Vion:2002}. This allows for simultaneous realization
of the maximum inductance (i.e., the maximum fluctuations of the phase
across the ladder) and the minimal coupling to the flux noise. This
regime corresponds to the optimal value of the ratio $E_{\textrm{JL}}/E_{\textrm{JS}}$,
which we denote as~$r_{\textrm{o}}$. The~$r_{\textrm{o}}$ value,
calculated within the classical approximation ($E_{\textrm{C}}=0$),
varies between 3 for a single unit cell and 5 for a very long ladder.
Quantum fluctuations renormalize $E_{\textrm{J}}(\varphi)$ and reduce
$r_{\textrm{o}}$ by $\delta r$; for our devices $\delta r\sim0.5$-0.7.

We have experimentally studied two types of ladders with the number
of unit cells $N=6$ and 24. Short ladders with $N=6$ allow for direct
comparison between experimental data and simulations based on the
numerical diagonalization of the system Hamiltonian. Because these
simulations are infeasible for a larger number of unit cells, the
ladders with $N=24$ were treated within the quasiclassical approximation.
These longer ladders demonstrate the potential of our novel approach:
their microwave impedance exceeds $R_{\textrm{Q}}$ by an order of
magnitude.

The ladders and the readout circuits were fabricated using multiangle
electron-beam deposition of Al films through a liftoff mask (see Supplemental
Material IA for details). The in-plane dimensions of small and large
junctions were $0.16\times0.16\,\mu\textrm{m}^{2}$ and $0.30\times0.30\,\mu\textrm{m}^{2}$,
respectively. The unit cell size was $3\times5\,\mu\textrm{m}^{2}$,
and the flux $\Phi=\Phi_{0}/2$ was realized in the magnetic field
$B\approx0.7$\,G. The ladders parameters are listed in Table~\ref{tab:data}.
Several devices with systematically varied values of $r$ were fabricated
on the same chip and inductively coupled to the same microstrip line.
The devices could be individually addressed due to their different
resonance frequencies. 

\begin{figure}[b]
\includegraphics[width=1\columnwidth]{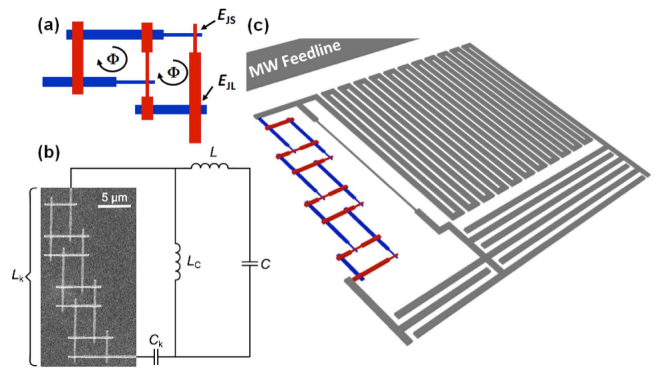} \caption{(color online) Schematic description of the on-chip circuit. (a)~Topology
of two unit cells of the ladder. Josephson contacts are formed at
the intersections of the bottom (horizontal) and top (vertical) electrodes.
(b)~The circuit diagram of the superinductor and $LC$ resonators
coupled via the kinetic inductance $L_{\textrm{C}}$ of a narrow superconducting
wire. The micrograph shows a ladder with six unit cells. (c)~The
on-chip circuit layout of two resonators inductively coupled to the
microwave (MW) feedline.}

\label{fig:circuit} 
\end{figure}

\begin{table*}[t]
\caption{Josephson junction parameters. $E_{\textrm{CS}}$ and $E_{\textrm{CL}}$
are the charging energies for the small and large junctions. }

\begin{tabular}{|c|c|c|c|c|c|c|c|c|c|c|}
\hline 
Device & Number & $E_{\textrm{JS}}$, & $E_{\textrm{CS}}$, & $E_{\textrm{JL}}$, & $E_{\textrm{CL}}$, & $r=$ & $C_{\textrm{K}}$, & $L_{\textrm{C}},$ & $L_{\textrm{K}}(\Phi=0)$, & $L_{\textrm{K}}(\Phi=\Phi_{0}/2)$,\tabularnewline
 & of unit cells & K & K & K & K & $E_{\textrm{JL}}/E_{\textrm{JS}}$ & fF & nH & nH & nH\tabularnewline
\hline 
\hline 
1 & 6 & 3.2 & 0.46 & 14.5 & 0.15 & 4.5 & 18 & 0.4 & 3.7 & 130\tabularnewline
\hline 
2 & 24 & 3.15 & 0.46 & 14.5 & 0.15 & 4.6 & 5 & 0.8 & 16 & 3\,000\tabularnewline
\hline 
\end{tabular}

\label{tab:data} 
\end{table*}

The effective kinetic inductance of the ladder, $L{}_{\textrm{K}}=(\omega_{01}^{2}C_{\textrm{K}})^{-1}$,
was calculated from the measured frequency~$\omega_{01}$ of the
$|0\rangle\leftrightarrow|1\rangle$ transition in the resonance circuit
formed by the ladder and the interdigitated capacitor. The capacitance
$C_{\textrm{K}}$ is larger than the capacitance of the interdigitated
capacitor due to the parasitic capacitance to the ground. The resonance
frequency of this circuit, referred below as the superinductor resonator,
varies with the magnetic field by an order of magnitude (see below),
whereas the bandwidth of the cryogenic preamplifier and cold circulators
in our measuring setup is limited to the range of 3-10\,GHz (the
microwave setup is described in Supplemental Material IB). To overcome
this limitation, the superinductor resonator was coupled to a linear
lumped-element $LC$ resonator with a resonance frequency $\omega_{LC}/2\pi\sim7$~GHz
via the coupling inductor $L_{\textrm{C}}$ {[}a narrow superconducting
wire, Figs.~\ref{fig:circuit}(b) and \ref{fig:circuit}(c){]}. The
$LC$ resonator is formed by an inductor (meandered 2-$\mu$m-wide
Al wire) with $L\lyxmathsym{ }=\lyxmathsym{ }5$\,nH and an interdigitated
capacitor (2-$\mu$m-wide fingers with 2\,$\mu$m spacing between
them) with $C\lyxmathsym{ }=\lyxmathsym{ }100$\,fF {[}Fig.~\ref{fig:circuit}(c){]}.
Both the superinductor and $LC$ resonators are inductively coupled
to a 2-port Al microstrip feedline with a 50\,$\Omega$ wave impedance.
When the superinductor resonator is excited by a second-tone microwave
frequency $\omega_{2}$, its impedance changes due to nonlinearity;
this results in a shift of the resonance of the $LC$ resonator probed
by the first-tone microwaves with $\omega_{1}\approx\omega_{LC}$.
The maximum shift occurs at the frequency $\omega_{2}=\omega_{01}$
that corresponds to the $|0\rangle\leftrightarrow|1\rangle$ transition.

\begin{figure}[b]
\includegraphics[width=0.8\linewidth]{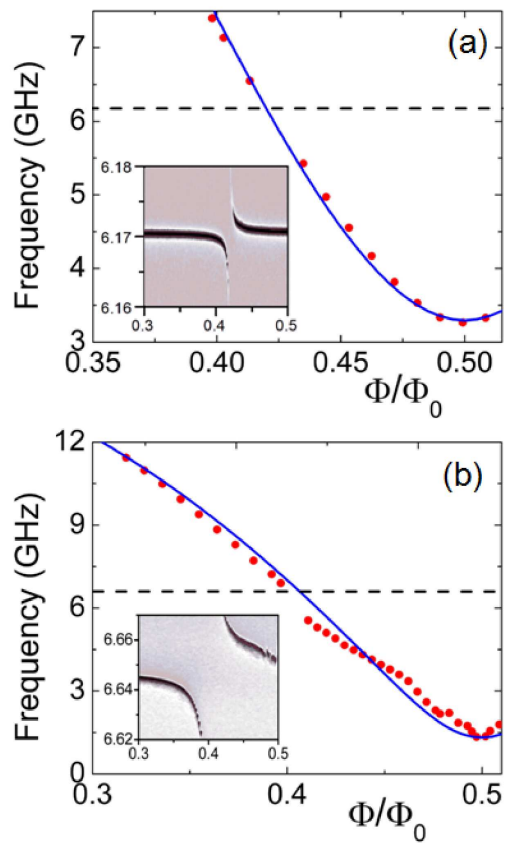} \caption{(color online) Spectroscopic data for the ladders with $N=6$ (a)
and $N=24$ (b). The resonance frequencies $\omega_{01}(\Phi)/2\pi$
of the $|0\rangle\leftrightarrow|1\rangle$ transition measured in
the second-tone experiments are shown by red dots. The horizontal
dashed lines correspond to the resonance frequency of the $LC$ resonators.
The blue curves represent the fits based on numerical diagonalization
of the circuit Hamiltonian for device~1 (a)~and the quasiclassical
modeling for device~2 (a); for the simulation parameters, see Table~\ref{tab:data}.
The gray scale insets show the microwave amplitude $|S_{21}|$ versus
the first-tone microwave frequency $\omega_{1}/2\pi$ and the normalized
flux $\Phi/\Phi_{0}$ measured near the avoided level crossings. }

\label{fig:spectroscopic} 
\end{figure}

Figure~\ref{fig:spectroscopic} shows the resonance modes corresponding
to the transition $|0\rangle\leftrightarrow|1\rangle$ in the superinductor
resonators with the 6-cell and 24-cell ladders. The insets in Fig.~\ref{fig:spectroscopic}
show the avoided crossing between the lowest modes of the superinductor
and $LC$ resonators observed in the first-tone measurements (these
avoided crossings illustrate the strength of coupling between these
resonators). The measured values of the lowest-mode frequency $\omega_{01}$
for the 6-cell ladder are in excellent agreement with simulations
based on the numerical diagonalization of the circuit Hamiltonian
(see Supplemental Material IIB). The only fitting parameter in these
calculations was the ratio $r=E_{\textrm{JL}}/E_{\textrm{JS}}$, which
is within 15\,\% of the designed value. The nominal junction parameters
for the 24-cell ladder also agree with the quasiclassical simulation
of the dependence $\omega_{01}(\Phi)$.

Superinductor applications require that the frequency~$\omega_{01}$,
which corresponds to the \textquotedblleft{}global\textquotedblright{}
mode of the superinductor, should be much smaller than the frequency
of its internal excitations, $\omega_{\textrm{int}}$. Far away from
full frustration, the internal modes correspond to very high frequencies
of an order of the Josephson plasma frequency ($\sim100$\,GHz for
our samples). At full frustration,~$\omega_{\textrm{int}}$ decreases,
but, according to our estimate, remains above 10\,GHz for the 24-cell
device. 

The inductance at full frustration increases as $r$ approaching its
optimal value $r_{\textrm{o}}$, which depends on the number of unit
cells (see Supplemental Material IIB): $r_{\textrm{o}}=4.1$ for $N=6$
and $r_{\textrm{o}}=4.5$ for $N=24$. Even though both devices 1
and 2 have similar values of $r$ (i.e. nominally identical junction
parameters), their proximity to the critical point is significantly
different. For device~1 with $r-r_{\textrm{o}}\approx0.4$, the inductance
at full frustration exceeds that at zero field by a factor of 35.
For device~2 with $r-r_{\textrm{o}}\approx0.1$, this increase exceeds
two orders of magnitude, and the inductance at full frustration is
3\,$\mu$H (for comparison, this is the inductance of a 3-meter-long
wire). The total capacitance of the superinductor resonator, $C_{\textrm{K}}=5$\,fF
(see Table~\ref{tab:data}), includes the capacitance of the interdigitated
capacitor (2\,fF) and the parasitic capacitance of all wires and
the superinductor to the ground (3\,fF) obtained by circuit modeling.
By \textquotedblleft{}meandering\textquotedblright{} the ladder and
shrinking the tested moderately-sized ($3\times5\,\mu\textrm{m}^{2}$)
unit cell, the parasitic capacitance can be reduced down to $\sim\lyxmathsym{ }1$\,fF;
the impedance of such a ladder approaches 50\,k$\Omega$ at $\omega/2\pi=\lyxmathsym{ }3$\,GHz. 

The nonlinearity of the superinductor-based qubit increases dramatically
with approaching the optimal working regime ($r=r_{\textrm{o}}$ and
$\Phi=\Phi_{0}/2$). According to our quantum simulations, the nonlinearity
factor $\gamma=(\omega_{12}-\omega_{01})/\omega_{01}$ ($\omega_{12}$
is the frequency of the $|1\rangle\leftrightarrow|2\rangle$ transition)
for the 24-cell ladder (device~2) with $r-r_{\textrm{o}}\approx0.1$
approaches 40\,\% at full frustration. Strong, tunable quartic nonlinearity
of the studied superinductor is an asset for the qubit design. In
particular, strong anharmonicity enables fast qubit operations and
qubit readout due to a large energy difference between the $|0\rangle\leftrightarrow|1\rangle$
and $|1\rangle\leftrightarrow|2\rangle$ transitions \cite{Zorin:2009}.

\begin{figure}[t]
\includegraphics[width=0.8\columnwidth]{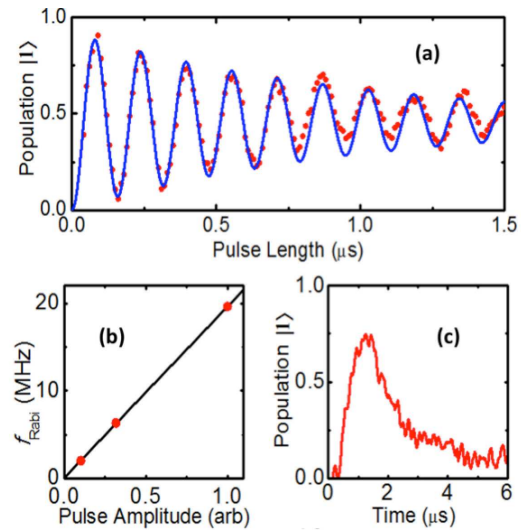} \caption{(color online) (a)~Rabi oscillations of the population of the first
excited level of the superinductor resonator in device~1. The phase
shift of the $LC$ resonance was measured while the superinductor
resonator was excited by the second-tone pulsed microwaves with $\omega_{2}=\omega_{01}$.
The data are shown for the phase $\varphi\lyxmathsym{ }=2\pi\Phi/\Phi_{0}=0.94\pi$.
The solid line represents the fit with the Rabi decay time 1.4\,$\mu$s.
(b)~Dependence of the Rabi frequency on the amplitude of microwaves
with $\omega_{2}=\omega_{01}$. (c)~The response of the $LC$ resonator
(measured at $\omega_{1}=\omega_{LC}$) to the excitation of the superinductor
resonator by a 0.4\,$\mu$s second-tone pulse. }

\label{fig:Rabi} 
\end{figure}

In order to demonstrate the high quality of our superinductor, we
have measured Rabi oscillations in the qubit formed by the superinductor
and the capacitor $C_{\textrm{K}}$ {[}Fig.~\ref{fig:Rabi}(a){]}.
In these measurements, we have monitored the phase shift of the $LC$
resonance while the superinductor resonator was excited by pulsed
microwaves with $\omega_{2}=\omega_{01}$. The quantum nature of these
oscillations was verified by observing the linear dependence of the
Rabi frequency on the amplitude of the microwave field {[}Fig.~\ref{fig:Rabi}(b){]}.
The observed decay time of Rabi oscillations exceeded 1\,$\mu$s
and was limited by the energy relaxation time {[}cf. Fig.~\ref{fig:Rabi}(c){]}.
The dominant source of energy relaxation is the intentional inductive
coupling to the $LC$ resonator and the microwave feedline.

The intrinsic decoherence rate of the qubit is expected to be very
low. Because the curvature $\partial^{2}E_{\textrm{J}}(\varphi)/\partial\varphi^{2}$
(which controls the position of energy levels) has a minimum at full
frustration, one expects that the flux noise does not affect the qubit
in the linear order. Another common source of dephasing in a chain
of superconducting islands coupled by Josephson junctions is the phase-slip
processes in combination with ubiquitous fluctuations of offset charges
on the islands \cite{Manucharyan:2012}. Because of the Aharonov-Casher
effect \cite{Manucharyan:2009,Elion:1993,Pop:2012,Koch:2009}, these
two factors result in the decoherence which is directly proportional
to the phase-slip rate. In the studied devices this rate, being proportional
to $\exp[-c(E_{\textrm{JL}}/E_{\textrm{CL}})^{1/2}]$, where $c\approx2.5$-2.8,
is expected to be negligible due to a large ratio $E_{\textrm{JL}}/E_{\textrm{CL}}$
($\approx100$) for the junctions that form the ladder backbone {[}Fig.~\ref{fig:setup}(a){]}.
Recent work~\cite{Masluk:2012} demonstrated that linear chains of
Josephson junctions with $E_{\textrm{J}}/E_{\textrm{C}}\approx100$
are phase-slip-free and exhibit inductances up to 0.3\,$\mu$H.

We envision many applications for the designed superinductor: this
element has the potential to reduce the charge noise sensitivity of
Josephson qubits, enable implementation of the fault tolerant qubits,
and provide sufficient isolation for the electrical current standards
based on Bloch oscillations. The ability to transform this element
from an inductor with an almost linear response into a very nonlinear
superinductor by tuning the magnetic field can facilitate controllable
coupling between qubits. In a moderately or strongly nonlinear regimes,
the superinductor-based resonator can also operate as a qubit with
a low decoherence rate. Being combined with a small Josephson junction,
the superinductor can be used as an adiabatic switch~--- an element
whose impedance changes exponentially with magnetic field \cite{Kitaev:2006,Matveev:2002},
which is crucial for the fault tolerant qubit operations \cite{Doucot:2012}.

We would like to thank V.~Manucharyan and A.~Zamolodchikov for helpful
discussions. The work was supported by DARPA (HR0011-09-1-0009), NSF
(DMR 1006265), and ARO (W911NF-09-1-0395).

\appendix

\section{Experimental details}

\subsection{Device fabrication}

The superinductor, the lumped-element $LC$ resonator, and the microstrip
feedline line were fabricated within the same vacuum cycle using multi-angle
electron-beam deposition of Al films through a nanoscale lift-off
mask. To minimize the spread of the junction parameters, we have used
the so-called \textquotedblleft{}Manhattan-pattern\textquotedblright{}
bi-layer lift-off mask formed by a 400-nm-thick e-beam resist (the
top layer) and 50-nm-thick copolymer (the bottom layer). In this technique,
the circuit is formed by aluminum strips of a well-controlled width
intersecting at right angles (thus, the \textquotedblleft{}streets\textquotedblright{}
and \textquotedblleft{}avenues\textquotedblright{}). After depositing
the photoresist on an undoped Si substrate and exposing the pattern
with e-beam lithography, the sample was placed in an ozone asher to
remove any traces of the photoresist residue. This step is crucial
for reducing the spread in junction parameters. 

The substrate was then placed in an oil-free high-vacuum chamber with
a base pressure of $5\times10^{-9}$\,mbar. The rotatable substrate
holder is positioned at an angle of $45^{\circ}$ with respect to
the direction of e-gun deposition of Al. During the first Al deposition,
the substrate was positioned such that the \textquotedblleft{}streets\textquotedblright{}
were orthogonal to the deposition direction. If the width of lines
in the e-beam resist forming the \textquotedblleft{}streets\textquotedblright{}
is less than the mask thickness ($0.45\,\mu\textrm{m}$), no metal
is deposited in the \textquotedblleft{}streets\textquotedblright{}
during this deposition, whereas the \textquotedblleft{}avenues\textquotedblright{}
are fully covered with Al. Alternatively, if the lines forming the
\textquotedblleft{}streets\textquotedblright{} are wider than $0.5\,\mu\textrm{m}$,
both the \textquotedblleft{}streets\textquotedblright{} and \textquotedblleft{}avenues\textquotedblright{}
are covered with metal in the first deposition. The lines narrower
than $0.5\,\mu\textrm{m}$ were used to form Josephson junctions,
the wider lines were used for the fabrication of meander-shaped inductors.
Without breaking vacuum, the surface of the bottom Al layer with a
thickness of 20\,nm was oxidized at $\thicksim100$\,mTorr of dry
oxygen for 5 minutes. 

After evacuating oxygen, the substrate holder was rotated by $90^{\circ}$,
and the second 60-nm-thick Al film was deposited; this layer forms
top electrodes of Josephson junctions, and increases the total thickness
of the meander-shaped inductors and interdigitated capacitors. Finally,
the sample was removed from the vacuum chamber and the lift-off mask
was dissolved in the resist remover. The spread of the resistances
for the nominally identical JJs with an area of $0.15\times0.15\,\mu\textrm{m}^{2}$
did not exceed 10\,\%.

\subsection{Measurement technique}

\begin{figure}[tb]
\includegraphics[height=4.1in]{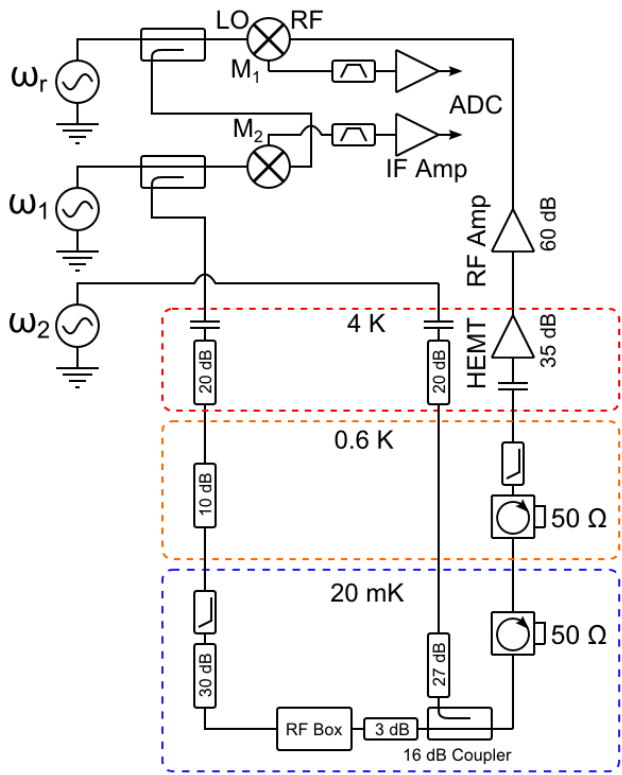} \caption{Simplified circuit diagram of the measurement setup. The microwaves
at the probe frequency $\omega_{1}$ are transmitted through the microstrip
line coupled to the superinductor and $LC$ resonators. The microwaves
at $\omega_{r}$, after mixing with the microwaves at $\omega_{1}$,
provided the reference phase $\phi_{0}$. The signal at $\omega_{1}$
is amplified, mixed down to an intermediate frequency $\omega_{1}-\omega_{r}$
by mixer M$_{2}$, and digitized by a fast digitizer (ADC). The second
channel of the ADC is used to digitize the signal from mixer M$_{2}$.
The microwaves at the second-tone frequency $\omega_{\textrm{2}}$
were coupled to the transmission line via a 16\,dB coupler.}

\label{fig:measurement-setup} 
\end{figure}

The microwave response of the coupled superinductor and $LC$ resonators
was probed by measuring both the phase and amplitude of the microwaves
traveling along a microstrip feedline coupled to the resonators. A
simplified schematic of the microwave circuit is shown in Fig.~\ref{fig:measurement-setup}.
The cold attenuators and low-pass filters in the input microwave line
prevented leakage of thermal radiation into the resonator. On the
output line, two cryogenic Pamtech isolators ($\sim18$\,dB isolation
between 3 and 12\,GHz) anchored to the mixing chamber were used to
attenuate the $\sim5$\,K noise from the cryogenic amplifier.

The probe microwaves at frequency $\omega_{1}$, generated by a microwave
synthesizer (Anritsu MG3694B), were coupled to the cryostat input
line through a 16\,dB coupler. These microwaves, after passing the
sample, were amplified by a cryogenic HEMT amplifier (Caltech CITCRYO
1-12, 35\,dB gain between 1 and 12\,GHz) and two 30\,dB room-temperature
amplifiers. The amplified signal was mixed by mixer M$_{1}$ with
the local oscillator signal at frequency~$\omega_{r}$, generated
by another synthesizer (Gigatronics 910). The intermediate-frequency
signal $A(t)=A\sin(\Omega t+\phi_{0})+A_{\textrm{N}}(t)$ (where $A_{\textrm{N}}(t)$
is the noise term) at $\Omega=(\omega_{1}-\omega_{r})/2\pi=30\,\textrm{MHz}$
was digitized by a 1\,GS/s digitizing card (AlazarTech ATS9870).
The signal was digitally multiplied by $\sin(\Omega t)$ and $\cos(\Omega t)$
averaged over $\thicksim10^{6}$ periods, and its amplitude $A$ (proportional
to the microwave amplitude $|S_{21}|$) and phase $\phi$ were extracted
as $A=\langle[A(t)\sin(\Omega t)]^{2}+[A(t)\cos(\Omega t)]^{2}\rangle^{1/2}$
and $\phi=\arctan\{\langle[A(t)\sin(\Omega t)]^{2}\rangle/\langle[A(t)\cos(\Omega t)]^{2}\rangle\}$,
respectively. The reference phase $\phi_{0}$ (which randomly changes
when both $\omega_{1}$ and $\omega_{2}$ are varied in measurements)
was found using similar processing of the low-noise signal provided
by mixer M$_{2}$ and digitized by the second channel of the ADC.
This setup enables accurate measurements of small changes $\phi-\phi_{0}$
unaffected by the phase jitter between the two synthesizers. The low
noise of this setup allowed us to perform measurements at a microwave
excitation level of $-133$\,dBm which corresponded to a sub-single-photon
population of the tank circuit. In the second-tone measurements, the
superinductor resonator was excited by the microwaves at frequency
$\omega_{\textrm{2}}$ propagating along the same microwave feedline
that was used for the microwave transmission at $\omega_{1}$.

\section{Theoretical analysis}

Here we explain in detail the nature of the large inductances of the
Josephson ladders studied in this work, the role played by the field
induced frustration and the numerical computations which results we
compared with the data in the main text. We begin our discussion with
the analysis of the classical (the Coulomb energy $E_{\textrm{C}}=0$)
model. We then discuss the computation that takes into account quantum
fluctuations induced by a non-zero~$E_{\textrm{C}}$.

\subsection{Classical analysis\label{sub:Classical-analysis}}

\begin{figure}
\includegraphics[width=1\linewidth]{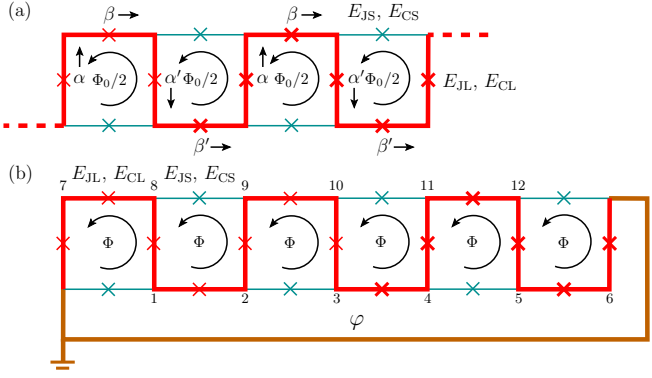} \caption{(a)~A small portion of a long fully frustrated Josephson ladder.
Large junctions with the Josephson energy $E_{{\rm JL}}$ and the
Coulomb energy $E_{{\rm CL}}\ll E_{\textrm{JL}}$ are shown in red
thick lines; they form the ladder backbone. Small junctions with the
Josephson energy $E_{{\rm JS}}$ and the Coulomb energy $E_{{\rm CS}}\ll E_{\textrm{JS}}$
are shown in cyan thin lines. The frustration of each cell is controlled
by the dimensionless parameter $\Phi/\Phi_{0}$. (b) Schematic diagram
of a six cell ladder (device 1). The ends of the ladder are grounded:
the phase difference across the ladder is $\varphi$. }

\label{fig:setup_ladder} 
\end{figure}

The main idea of the superinductor design becomes more transparent
in the case of a fully frustrated ($\Phi=\Phi_{0}/2$) long ladder
shown in Fig.~\ref{fig:setup_ladder}(a). As we explain below, the
fully frustrated regime is also very important experimentally because
in this regime the quantum states of the ladder are less sensitive
to the flux noise. 

It is convenient to choose the gauge in which the magnetic field induces
the phases $2\pi\Phi/\Phi_{0}$ on smaller junctions. Because at full
frustration these phases are equal to $\pi$, the shift of the ladder
by one unit cell combined with its rotation about its axis transforms
the ladder into itself, so the problem becomes translationally invariant.
Furthermore, a large length of the ladder allows to neglect the boundary
effects and focus on translationally invariant solutions. 

\begin{figure}[tb]
\includegraphics[height=1.45in]{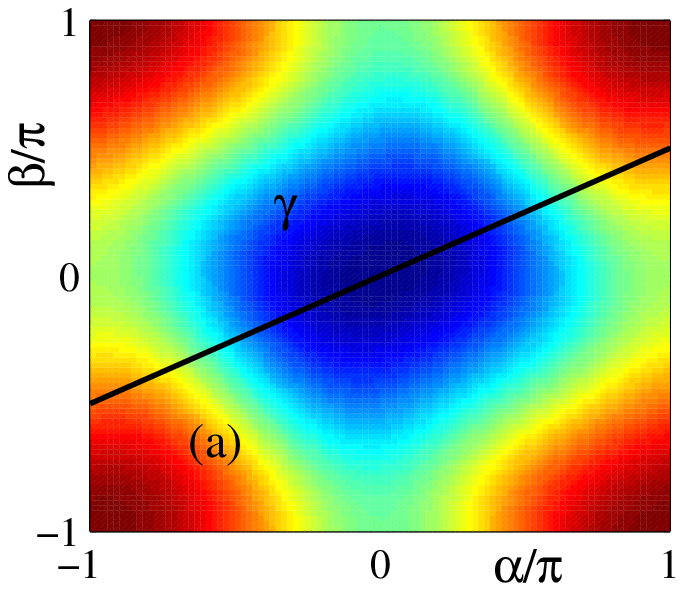} \includegraphics[height=1.45in]{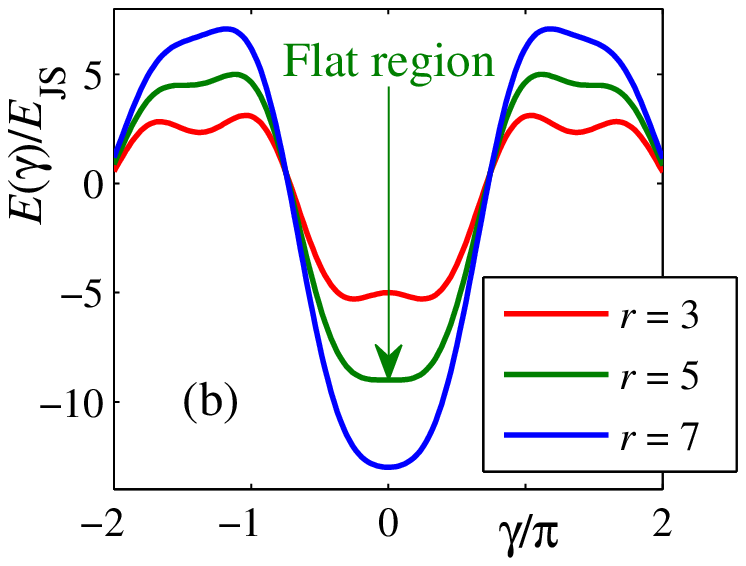}\includegraphics[height=1.45in]{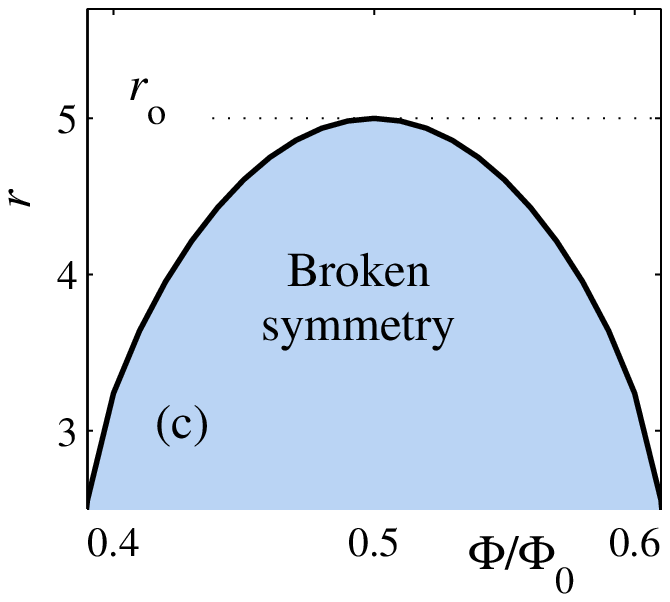}
\caption{(a)~Energy per elementary cell $E(\alpha,\beta)$ (\ref{eq:E(alpha,beta)})
as a function of $\alpha$ and $\beta$ at the optimal point $r=r_{\textrm{o}}$.
The black line corresponds to the flat direction parametrized by $\gamma$.
(b)~The dependence of energy in the flat direction for different
values of~$r$. (c)~Phase diagram of the model in classical approximation
in dimensionless variables $r=E_{\textrm{JL}}/E_{\textrm{JS}}$ and
$\Phi/\Phi_{0}$. Quantum fluctuations discussed in section \ref{sub:Quantum-case}
shift the transition line down but affect neither the long range order
nor the phase transition. }

\label{fig:EClassical} 
\end{figure}

Consider a small portion of the fully frustrated ladder shown in Fig.
\ref{fig:setup_ladder}(a). The translational invariance implies that
the solution is described by two phases across large junctions: $\alpha$
at the ``vertical'' junctions and $\beta$ at the ``horizontal''
ones. The total phase across the whole ladder is $\varphi=N(\alpha+\beta)$,
where $N$ is the number of rungs. The energy per unit cell that contains
two large and one small junctions is given by 
\begin{equation}
E(\alpha,\beta)=-E_{{\rm JL}}(\cos\alpha+\cos\beta)-E_{{\rm JS}}\cos(\pi-2\alpha-\beta).\label{eq:E(alpha,beta)}
\end{equation}
 The dependence $E(\alpha,\beta)$ is shown in Fig.~\ref{fig:EClassical}(a).
At $E_{\textrm{JL}}\gg E_{\textrm{JS}}$ the first term in Eq.~(\ref{eq:E(alpha,beta)})
dominates and $E(\alpha,\beta)$ has a minimum at $(\alpha,\beta)=(0,0)$.
The expansion near this point gives 
\begin{equation}
E^{(2)}(\alpha,\beta)=\frac{1}{2}\,[\begin{array}{cc}
\alpha & \beta\end{array}]\left[\begin{array}{cc}
E_{{\rm JL}}-4E_{{\rm JS}} & -2E_{{\rm JS}}\\
-2E_{{\rm JS}} & E_{{\rm JL}}-E_{{\rm JS}}
\end{array}\right]\left[\begin{array}{c}
\alpha\\
\beta
\end{array}\right],\label{eq: E^(2)(alpha,beta)}
\end{equation}
which is a quadratic form with eigenvalues $E_{{\rm JL}}$ and $E_{{\rm JL}}-5E_{{\rm JS}}$.
The second eigenvalue changes sign as the ratio $r=E_{\textrm{JL}}/E_{\textrm{JS}}$
decreases, signaling the instability and appearance of a nontrivial
ground state. The critical value of the ratio, $r_{\textrm{o}}=5$,
corresponds to the optimal working point, as explained below. At this
point the function $E(\alpha,\beta)$ becomes flat along the eigenvector
$(2,1)$. We introduce new coordinates in the ``flat'' and ``steep''
directions: $\gamma=(2\alpha+\beta)/\sqrt{5}$, $\delta=(\alpha-2\beta)/\sqrt{5}.$
Figure~\ref{fig:EClassical}(b) shows the energy plotted along the
flat direction. Neglecting the phase deviations in the steep direction,
the total phase is related to $\gamma$ by $\varphi=3N/\sqrt{5}\gamma$.

At $r>r_{\textrm{o}}$ the ground state values of phases $\gamma$
and $\varphi$, $\gamma_{0}$, and $\varphi_{0}$, are zero. Deviations
of the total phase from $\varphi_{0}=0$ result in the quadratic increase
of the energy 
\[
E^{(2)}(\varphi)=\frac{5}{18N}E_{\textrm{JS}}(r-r_{\textrm{o}})\varphi^{2}.
\]
At the optimal point ($r=r_{\textrm{o}}$) these deviations vanish,
and the inductance, which is inversely proportinal to $d^{2}E/d\varphi^{2}$,
becomes infinite in the quadratic approxiamtion and is defined by
the non-vanishing quartic term

\[
E^{(4)}(\varphi)=\left.\frac{625-17r}{48600N^{3}}E_{\textrm{JS}}\varphi^{4}\right|_{r=r_{\textrm{o}}}=\frac{1}{90N^{3}}E_{\textrm{JS}}\varphi^{4}.
\]
At smaller $r<r_{\textrm{o}}$ the ground state corresponds to a non-zero
value of $|\gamma_{0}(r)|\propto\sqrt{r_{\textrm{o}}-r}$ that translates
into a non-zero $\varphi_{0}=\pm3N\gamma_{0}/\sqrt{5}$. The appearance
of a non-zero $\gamma_{0}$ is equivalent to a phase transition.

The translational invariance discussed above is violated away from
full frustration. At arbitrary frustration, the ladder remains symmetric
under the translation by two unit cells. Denoting the phases across
the large junctions in the first cell by $\alpha,\beta$ and in the
second cell by $\alpha',\beta'$ we obtain the classical energy per
two rungs 
\begin{align}
E(\Phi) & =-E_{{\rm JL}}\left[\cos\alpha+\cos\beta+\cos\alpha'+\cos\beta'\right]\nonumber \\
- & E_{{\rm JS}}\left[\cos(\frac{2\pi\Phi}{\Phi_{0}}-\alpha_{+}-\beta)+\cos(\frac{2\pi\Phi}{\Phi_{0}}+\alpha_{+}+\beta')\right],\label{eq:E(Phi)}
\end{align}
where $\alpha_{+}=\alpha+\alpha'$ . This energy does not change under
the combined transformation $\alpha\rightarrow\alpha'$, $\beta\rightarrow\beta'$,
and $\Phi\rightarrow-\Phi$, which correspond to the shift by one
unit cell and simultaneous change in the sign of the magnetic field.
The total phase difference along the ladder $\varphi=N(\alpha_{+}+\beta_{+})/2$
is invariant under this transformation. Time inversion invariance
implies that the ground state energy does not change under the simultaneous
change in sign in both $\Phi$ and $\varphi$, the symmetry under
the combined transformation implies that the ground state energy is
an even function of $\varphi$ at any field. This conclusion is based
only on the symmetry of the problem and thus remains valid when quantum
fluctuations are taken into accound (see section \ref{sub:Quantum-case}).

The energy (\ref{eq:E(alpha,beta)}) is minimal at $\alpha_{-}=\alpha'-\alpha=0$
and $\beta_{-}=(\beta'-\beta)/2\neq0$ that has to be determined from
the minimization of (\ref{eq:E(Phi)}). The minimum in $\alpha_{-,}\beta_{-}$
is always steep, so fluctuations in this direction can be neglected.
In this approximation the general expression for the energy (\ref{eq:E(Phi)})
is reduced to the form (\ref{eq:E(alpha,beta)}) with the effective
\[
E_{\textrm{JS}}^{\textrm{(eff)}}=E_{\textrm{JS}}\cos(2\pi x+\beta_{-}),
\]
where $x=\Phi/\Phi_{0}-1/2$ . 

At $r>r_{\textrm{o}}$ the ratio $E_{\textrm{JL}}/E_{\textrm{JS}}^{\textrm{(eff)}}$
that controls the instability is large: $E_{\textrm{JL}}/E_{\textrm{JS}}^{\textrm{(eff)}}>r>r_{\textrm{o}}$,
so the ground state corresponds to zero phases. At $r<r_{\textrm{o}}$
the instability occurs at non-zero $x_{\textrm{c}}$, defined by $E_{\textrm{JL}}/E_{\textrm{JS}}^{\textrm{(eff)}}(x_{c})=r_{\textrm{o}}$
{[}see Fig.~\ref{fig:EClassical}(c){]}. At the point of instability
we can determine the value of $\beta_{-}$ assuming that $\alpha_{+}=\beta_{+}=0$,
we get

\begin{equation}
\beta_{-}(x)=\arctan\left[\frac{\sin(2\pi x)}{r-\cos(2\pi x)}\right].\label{eq:beta_-}
\end{equation}

The effective reduction of $E_{\textrm{JS}}$ implies that at $r>r_{\textrm{o}}$
the transition from the single minimum at $\gamma=0$ to double minimum
at $\pm\gamma_{0}$ occurs as a function of the field at~$x_{\textrm{c}}$
where 
\begin{equation}
\cos[2\pi x_{\textrm{c}}+\beta_{-}(x_{\textrm{c}})]=\frac{r}{r_{\textrm{o}}}.\label{eq:x_c}
\end{equation}
The transition line defined by Eqs.~(\ref{eq:beta_-}) and (\ref{eq:x_c})
separates a symmetry-broken (ordered) phase from the phase with unbroken
symmetry. We show the phase diagram obtained by the numerical solution
of these equations in Fig. \ref{fig:EClassical}(c). The equation
for the instability can be solved analytically for $r_{\textrm{o}}-r\ll r_{\textrm{o}}$
(i.e. $x\ll1$):
\[
x_{\textrm{c}}=\frac{r}{2\pi(r+1)}\arccos\left(\frac{r}{r_{\textrm{o}}}\right).
\]
At the critical flux $\Phi_{\textrm{c}}=\left(1/2\pm x_{\textrm{c}}\right)\Phi_{0}$
the quadratic part of the energy vanishes which implies infinite inductance.
Thus, the system can be tuned to the infinite inductance either by
realizing $r=r_{\textrm{o}}$ at full frustration, or by approaching~$\Phi_{\textrm{c}}$
at $r<r_{\textrm{o}}.$

In the absence of noises and disorder, these two methods of realization
of the infinite inductance (and the phase transition) are equivalent.
However, the real systems are subject to flux noise. Because the energy
of the ground state is quadratic in flux deviations $\delta\Phi$,
the effect of the flux noise on the system is minimized if the infinite
inductance is realized at full frustration, or close to it. Thus,
the optimal working point corresponds to the ratio $E_{\textrm{JL}}/E_{\textrm{JS}}=r_{\textrm{o}}$. 

\begin{figure}[tb]
\includegraphics[height=1.5in]{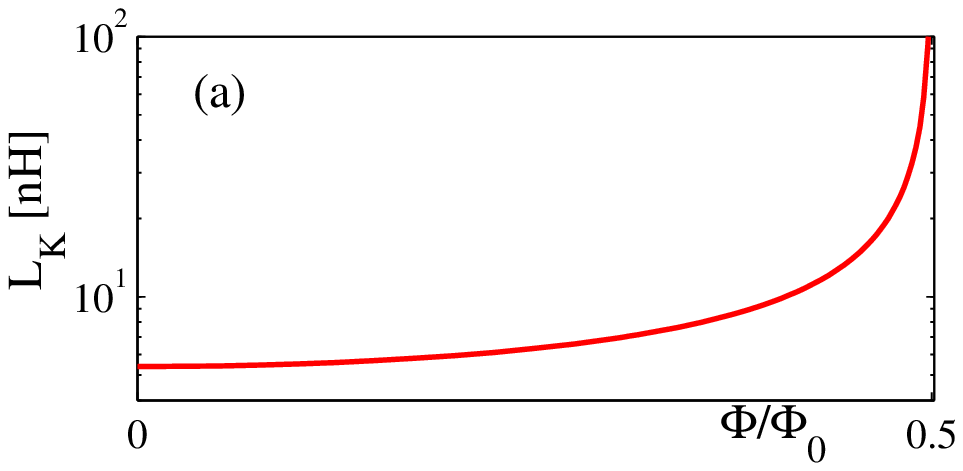} \includegraphics[height=1.5in]{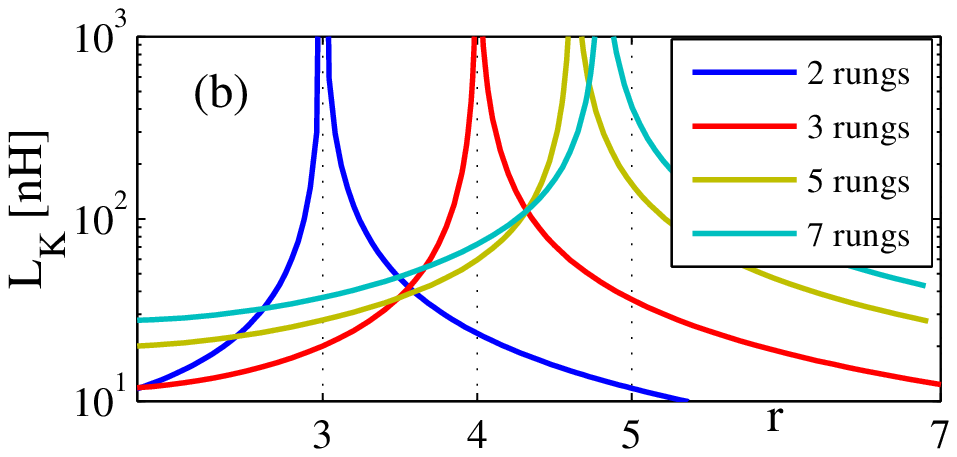}
\caption{(a)~Inductance as a function of frustration (external magnetic field)
in the classical approximation for a three-rung ladder at $r=r_{\textrm{o}}$.
(b)~The inductance as a function of the ratio $r=E_{{\rm JL}}/E_{{\rm JS}}$
for ladder with 2, 3, 5, and 7 rungs. Optimal ratios $r_{\textrm{o}}$:
$3$ for 2 rungs, $4$ for 3 rungs, $4.61$ for 5 rungs, and $4.80$
for 7 rungs.}

\label{fig:inductance_classical} 
\end{figure}

The finite length of the ladder brings in two effects associated with
the boundaries. First, this may violate the symmetry which resulted
in $E(\varphi)=E(-\varphi)$ for all frustrations; this is crucial
for the realization of infinite inductance at $\Phi_{c}\neq\Phi_{0}/2$.
However, this effect is absent for the odd number of rungs (this was
the reason for selecting the even number of unit cells in the studied
ladders). In this case the ladder remains symmetric under the rotation
by $\pi$ about its center. The rotation does not change the magnetic
fluxes but changes $\varphi\rightarrow-\varphi$, so in this case
the function $E(\varphi)$ remains even for all fluxes and $r$. Second,
the finite-size effects decrease the value of the critical $r$ at
which the instability occurs at full frustration. To compute them
for finite ladders we numerically computed the energy $E(\varphi)$
for the ladders containing from 2 to 7 rungs. The dependences of the
inductance $L_{\textrm{K}}\left(\Phi,r\right)=\left.\left(\partial^{2}E/\partial\varphi^{2}\right)^{-1}\right|_{\varphi=\varphi_{0}}$
for different ladder lengths are shown in Fig.~\ref{fig:inductance_classical}.
The value of $\varphi_{0}$is defined by the minimum of the ground
state energy $E(\varphi)$ as a function of the phase $\varphi$.
The inductance increases with frustration and diverges at full frustration
$\Phi=\Phi_{0}/2$ for $r=r_{\textrm{o}}$. For the ladders with six
unit cells, the finite size effects lead to a small decrease of the
optimal value to $r_{\textrm{o}}=4.8.$

\begin{figure}[tb]
\includegraphics[height=1.5in]{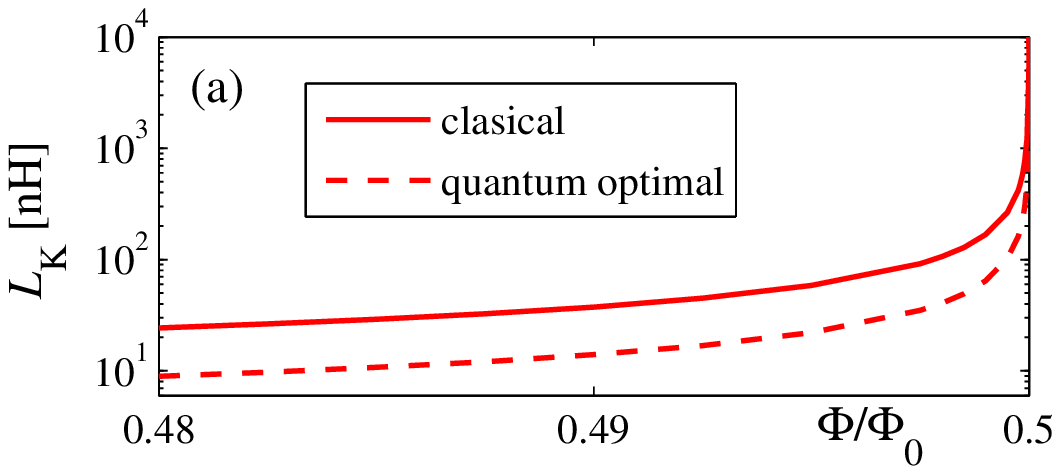} \includegraphics[height=1.5in]{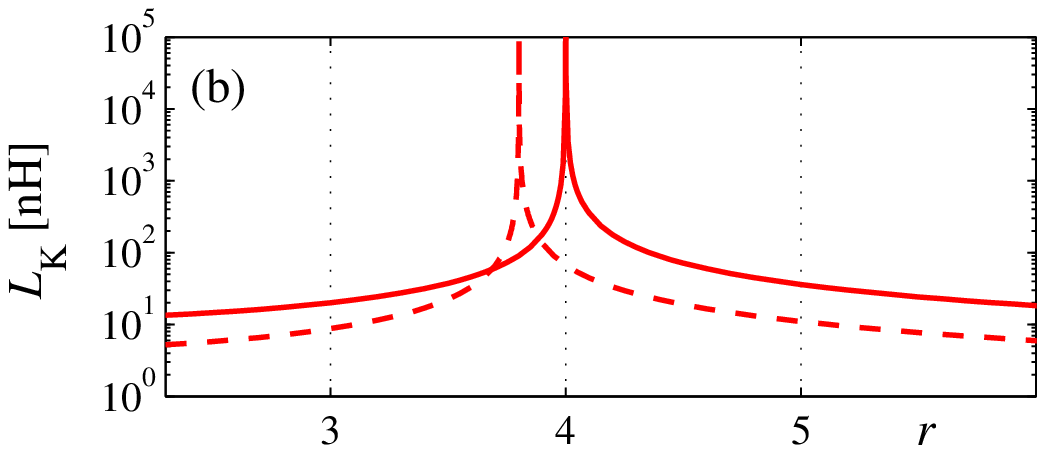}
\caption{Solid lines in panels (a) and (b) show the inductance computed in
the classical approximation for a three-rung ladder as a function
of frustration at $r=r_{\textrm{o}}$ (a) and as a function of $r$
for $\Phi=\Phi_{0}/2$ (b). Dashed lines show the results of the full
quantum computation for the same ladder. Quantum fluctuations shift
position of the optimal point from its classical value ($r_{\textrm{o}}=4$)
to $r_{\textrm{o}}\approx3.8$. Dashed line in panel (a) shows the
result of the full quantum computation at $r=r_{\textrm{o}}$ which
takes into account the shift of $r_{\textrm{o}}$ due to quantum fluctuations.}

\label{fig:inductance_quantum} 
\end{figure}

\subsection{Quantum analysis\label{sub:Quantum-case}}

Non-zero (but small) charging energies result in quantum fluctuations
of the phase of individual islands, $\theta_{i}$. These fluctuations
have three effects. First, they might result in a phase slip in which
the phase difference $\eta_{ij}=\theta_{i}-\theta_{j}$ between two
islands connected by a large Josephson junction changes by $2\pi$
due to a quantum tunneling process. Second, they may result in renormalization
of the effective energy in the flat direction discussed in section
\ref{sub:Classical-analysis} and the shift of the phase boundary
in Fig.~\ref{fig:EClassical}(c). Finally, they may have a very large
effect on the properties of long ladders close to the transition point
at which critical fluctuations are expected to become important.

In the absence of frustration the amplitude of a phase slip across
a single large junction, $E_{\textrm{ph}}\approx3.8E_{\textrm{CL}}^{1/4}E_{\textrm{JL}}^{3/4}\exp(-\sqrt{8E_{\textrm{JL}}/E_{\textrm{CL}}})$,
is exponentially small for the experimental values of $E_{{\rm JL}}/E_{{\rm CL}}\approx100$.
The frustration decreases the tunnel barrier resulting in larger phase
slip amplitudes. However, this effect is numerically small. Namely,
the dominant transitions correspond to the process in which the phase
$\beta$ changes between $0$ and $2\pi$ {[}see Fig. \ref{fig:EClassical}(a){]},
the action for this process is $S\approx2.5\sqrt{E_{\textrm{JL}}/E_{\textrm{CL}}}$.
Thus, the phase slip amplitudes remain negligible for the ladders
with experimental parameters.

Small quantum fluctuations of the phase differences smear the classical
potential. As a result the large inductance and the phase transition
are realized at a smaller value of $r_{\textrm{o}}.$ To estimate
this effect we have diagonalized numerically the Hamiltonian describing
small ladders. This diagonalization is convenient to perform in the
charge basis in which the general Hamiltonian of the Josephson circuit
acquires the form

\begin{align}
\hat{H}= & -\sum_{\langle i,j\rangle}\frac{J_{ij}}{2}\bigl(e^{-iA_{ij}}\mathbf{b}_{i}^{+}\mathbf{b}_{j}^{-}+{\rm H.c.}\bigr)\nonumber \\
 & +(2e)^{2}\sum_{i,j}(C^{-1})_{ij}(n_{i}+q_{i})(n_{j}+q_{j}),\label{eq:HamRot}
\end{align}
where matrix $J_{ij}$ describes Josephson couplings between islands
and $C_{ij}$ is the capacitance matrix of the system. For large junctions
$J_{i,i+1}=E_{\textrm{JL}}$, $C_{i,i+1}=C_{\textrm{L}}$, and $E_{\textrm{CL}}=e^{2}/2C_{\textrm{L}}$;
for small junctions $J_{i,i+3}=E_{\textrm{JS}}$, $C_{i,i+3}=C_{\textrm{S}}$,
and $E_{\textrm{CS}}=e^{2}/2C_{\textrm{S}}$. The operators $\mathbf{b}^{\pm}$
increase (decrease) the number of Cooper pairs on each island: $\mathbf{b}_{i}^{\pm}=|n_{i}\pm1\rangle_{i}\langle n_{i}|$.
The junction matrix $J_{ij}$ is symmetric and its diagonal terms
are zeros. The capacitance matrix $C_{ij}$ is positively defined
with positive diagonal elements $C_{ii}$ and negative off-diagonal
elements $C_{ij}$, $i\neq j$, so that its inverse $(C^{-1})_{ij}$
is positively defined and its elements are all positive. The phases
$A_{ij}$ are induced by the magnetic field, the sum of $A$'s over
a closed loop $l$ of $n$ Josephson junctions corresponds to the
flux $\Phi_{l}$ penetrating this loop.

For the numerical diagonalization we had to limit the number of charging
states, $n_{q}$ at each island. For the parameters similar to the
experimental ones it is sufficient to keep $n_{q}=12$ to get the
results with better than $1\%$ accuracy. The effect of the quantum
fluctuations on the inductance of a three rung chain with $E_{{\rm JL}}/E_{{\rm CL}}=40$
is shown in Fig. \ref{fig:inductance_quantum}. We observe that quantum
fluctuations reduce $r_{\textrm{o}}$ by $\delta r\approx0.2$ down
from its classical value, but the behavior of the inductance as a
function of the frustration remains essentially the same. The importance
of quantum fluctuations increases with the ladder length, so this
shift becomes $\delta r\approx0.5$ for the ladders with six unit
cells.

The inductance $L$ can be calculated as the inverse curvature of
the ground state energy $E(\varphi)$. Alternatively, it can be defined
as $L=1/\omega_{01}^{2}C$ for a superinductor shunted by a capacitor
$C$. Here the frequency $\omega_{01}$ is the excitation frequency
of the superinductor resonator $\omega_{01}=E_{1}-E_{0}$. In the
limit of a large capacitance $C$ both methods give the same result
(for a linear circuit, the results of both methods are identical for
any $C$).

For the detailed comparison with the data for device~1 we have diagonalized
the Hamiltonian of the ladder connected to a large capacitor. In this
Hamiltonian we used the values of the charging energy of small junctions
and their Josephson energy extracted from the independent measurements
of the Cooper pair transistors containing nominally identical junctions
with $E_{\textrm{CS}}=0.46\mathrm{\,\textrm{K}}$ and $E_{\textrm{JS}}=3.2\mathrm{\,\textrm{K}}$.
The value of the charging energy of the large junction that was calculated
from the ratio of the areas of small and large junctions: $E_{\textrm{CL}}=0.15\mathrm{\,\textrm{K}}$.
We used the designed value of the large capacitance $C_{\textrm{K}}=18\,\mathrm{\textrm{fF}}$.
We then diagonalized the Hamiltonian of the ladders containing up
to six rungs, determined the excitation frequency of the superinductor
resonator, $\omega_{\textrm{01}}(\Phi)\equiv E_{1}-E_{0}$, and extrapolated
the result to 7 rungs that cannot be computed directly due to an enormous
size of the Hilbert space. For the experimental values of the charging
energies $r_{\textrm{o}}$ is shifted down to approximately $4.1$.
The 7-rung devices studied in this work (e.g. device 1) are characterized
by the ratios $r$ that are significantly above $r_{\textrm{o}}$,
so the extrapolation to seven rungs works reasonably well. We found
that in this parameter range the form of the dependence $\omega_{\textrm{01}}(\Phi)$
in the interval $0.4<\Phi/\Phi_{0}<0.5$ is very similar for different
$r$, whilst the absolute values of $\omega_{\textrm{01}}(0.4\Phi_{0})$
and $\omega_{\textrm{01}}(0.5\Phi)$ change dramatically as a function
of $r$. This dependence is plotted in Fig.~\ref{fig:spectroscopic}(a).
Fitting the ratio $\omega_{\textrm{01}}(0.4\Phi_{0})/\omega_{\textrm{01}}(0.5\Phi_{0})$
to experimental values for device~1 gives for the ratio $r=4.5$
that is close to the designed value of $4.0$.

\begin{figure}
\includegraphics[width=2.3in]{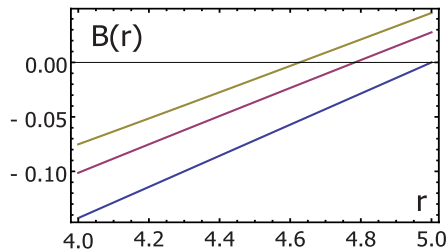}

\caption{Classical and quantum $B(r)$. From right to left: the results of
classical computation and from the full diagonalization of Hamiltonians
for three and four rungs ladders. }

\label{fig:B(r)}
\end{figure}

The energy of the first excited states of longer device~2 was obtained
in a two step procedure. In the first step we used exact diagonalization
to find the energy functional of small quantum ladders (with 3-5 rungs)
forming the loop penertrated by flux $\Phi=\Phi_{0}/2$ and fit it
to the general form expected in vicinity of the optimal point: 
\begin{equation}
E(\varphi)=\frac{B_{\textrm{q}}(r)}{N}E_{\textrm{JS}}\varphi^{2}+\frac{C_{\textrm{q}}(r)}{N^{3}}E_{\textrm{JS}}\varphi^{4}.\label{eq:Ephi}
\end{equation}
Here $B_{\textrm{q}}(r)$ and $C_{\textrm{q}}(r)$ are numerical coefficients
that differ somewhat from their classical values, $B_{\textrm{cl}}$
and $C_{\textrm{cl}}$. Comparison of the results of the computation
with classical result shows that the main effect of quantum computation
is a shift of the critical value of $r$ (see Fig.~\ref{fig:B(r)}),
$B_{\textrm{q}}(r)\approx B_{\textrm{cl}}(r+\delta r)$ with $\delta r\approx0.45$
and a modest renormalization of $C_{\textrm{q}}(r)\approx0.7C_{\textrm{cl}}(r=r_{\textrm{o}})$
that can be approximated by the constant in the relevant regime.

The procedure outlined above allowed us to determine the renormalization
of the classical energy by short scale quantum fluctuations. In the
second step we used Eq.~(\ref{eq:Ephi}) to compute the potential
energy of the long ladder of device 2. Solving finally the quantum
problem of the phase fluctuations across the whole ladder characterized
by the Hamiltonian
\[
H=E(\varphi)+4E_{\textrm{C}}q^{2}
\]
we find the low energy excitations of the device. The results are
shown in Fig.~\ref{fig:spectroscopic}(b).
\end{document}